\documentclass[aip,apl,reprint,showpacs,amsmath,amssymb,floatfix,citeautoscript]{revtex4-1}

\usepackage{graphicx}% Include figure files
\usepackage{dcolumn}% Align table columns on decimal point
\usepackage{bm}% bold math

\usepackage{xcolor}
\usepackage{SIunits}
\usepackage{comment}
\usepackage{ulem}

\begin{document}

\title{Tuning of the hole spin relaxation time in single self-assembled In$_{1-x}$Ga$_x$As/GaAs
quantum dots by electric field}

\author{Hai Wei}
\affiliation{Key Laboratory of Quantum Information, University of Science and Technology of China,
Hefei, Anhui, 230026, China}
\affiliation{Synergetic Innovation Center of Quantum Information and Quantum
  Physics, University of Science and Technology of China, Hefei, 230026, China}

\author{Guang-Can Guo}
\affiliation{Key Laboratory of Quantum Information, University of Science and Technology of China,
Hefei, Anhui, 230026, China}
\affiliation{Synergetic Innovation Center of Quantum Information and Quantum
  Physics, University of Science and Technology of China, Hefei, 230026, China}

\author{Lixin He}
\email{helx@ustc.edu.cn}
\affiliation{Key Laboratory of Quantum Information, University of Science and
Technology of China, Hefei, Anhui, 230026, China}
\affiliation{Synergetic Innovation Center of Quantum Information and Quantum
  Physics, University of Science and Technology of China, Hefei, 230026, China}

\date{\today}

\begin{abstract}
We investigate the electric field tuning of the phonon-assisted hole spin relaxation
in single self-assembled In$_{1-x}$Ga$_{x}$As/GaAs quantum dots, 
using an atomistic empirical pseudopotential method.
We find that the electric field along the growth direction can 
tune the hole spin relaxation time for more than 
one order of magnitude. The electric field can prolong or shorten the hole
spin lifetime and the tuning shows an asymmetry in terms of
the field direction. The asymmetry is more pronounced for 
the taller the dot. The results show that
the electric field is an effective way to tune the hole spin-relaxation in self-assembled QDs.
\end{abstract}

% PACS, the Physics and Astronomy
                             % Classification Scheme.
%\keywords{Suggested keywords}%Use showkeys class option if keyword
                              %display desired
\maketitle

Because of the three-dimensional confinement, the electron and hole in
self-assembled quantum dots (QDs) are only weakly coupled to the environment, 
and therefore have much longer spin lifetimes than their counterparts in bulk
materials.~\cite{meier84,woods02}
They have thus been proposed as the quantum bits (qubits) for quantum
information processes. \cite{loss98,kane98}
Recently, the initialization,
manipulation and readout of electron/hole spins in QDs have been demonstrated
experimentally.~\cite{kroutvar04,braun05,heiss07,gerardot08}

The hole spins are expected to have long coherence time, 
because the hyperfine interaction between hole spin 
with the nuclear spins is relatively small.~\cite{eble09}
The main mechanism that leads to the hole spin
relaxation is the spin-phonon interaction due
to spin-orbit coupling (SOC).~\cite{cheng04,heiss07,gerardot08,trif09,wei12,warburton13,bulaev05}
As we know, the Dresselhaus SOC originates from bulk inversion asymmetry
(BIA)~\cite{dresselhaus55} 
and Rashba SOC originates from structure inversion asymmetry
(SIA).~\cite{bychkov84}
Therefore, it is possible to tune the SOC by applying external fields, which may
change both BIA and SIA in the QDs.
Recent experimental~\cite{balocchi11,kanai11} and
theoretical~\cite{prabhakar12,prabhakar13} studies have shown that the SOC
strength can be enhanced by the in-plane electric and magnetic field indeed.
As a consequence, the hole spin relaxation can also be tuned by the external
electric field.

In this paper, we investigate the tuning 
of hole spin relaxation time ($T_1^h$)
by applying an external electric field along the QDs growth direction using an
atomistic empirical pseudopotential method (EPM).~\cite{williamson00} 
We find that the $T_1^h$ can be tuned by more than one order of magnitude
by the external field. It is therefore an effective way to tune the spin relaxation
in self-assembled QDs.

\begin{figure*}
\includegraphics[width=0.8\textwidth]{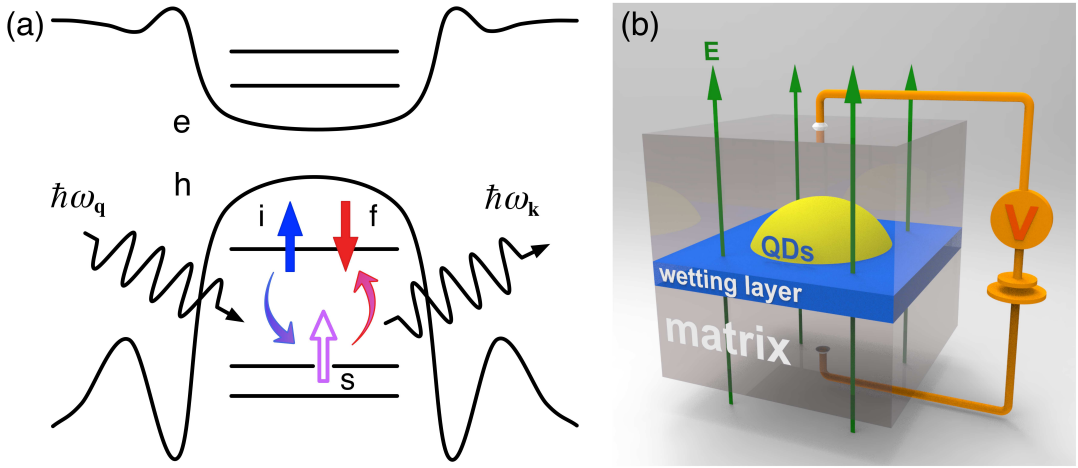}
% Here is how to import EPS art
\caption{(Color online) (a) The schematic show of the second-order
  phonon-assisted hole spin relaxation in self-assembled QDs. The $i$, $f$
  and $s$ are the initial, final and intermediate levels, respectively.
(b) The schematic drawing of single lens-shaped In$_{1-x}$Ga$_x$As/GaAs QDs
embedded in the GaAs matrix. The dot is grown on a monolayer wetting layer.
The external electric field is applied along the
[001] ($E>$0) or $[00\bar{1}]$ ($E<$0) direction.}
\label{fig:QDs}
\end{figure*}

We study the hole spin relaxation at low magnetic field ($B_z = 1$ mT), 
where the spin relaxation is dominant by the
two-phonon process.~\cite{wei12,trif09,fras12}
As schematically shown in Fig.~\ref{fig:QDs}(a), a hole at the initial
(labeled as $i$) state, with energy $\epsilon_i$, 
absorbs a phonon of momentum $\mathbf{q}$ and jumps to an intermediate
($s$) state with energy $\epsilon_s$. It then emits a phonon with momentum $\mathbf{k}$ and relaxes
to the final ($f$) state with energy $\epsilon_f$, which has an opposite spin of the initial state.
The hole spin-flip rate ($ \tau_{\nu}^{-1} $) from the initial to the final state is given by
the second-order Fermi's Golden Rule,~\cite{wei12}
\begin{eqnarray}
\frac{1}{\tau_{\nu}} & = &
\frac{2\pi}{\hbar}\sum_{\mathbf{q,k}}%\sum_{\mathbf{k}}
\left[ \sum_{s}'
(\frac{M_{\mathbf{q}}^{is}M_{\mathbf{k}}^{sf}}{\epsilon_{i}-\epsilon_{s}+\hbar \omega_{\mathbf{q}}} 
+\frac{M_{\mathbf{k}}^{is}M_{\mathbf{q}}^{sf}}{\epsilon_{i}-\epsilon_{s} - \hbar \omega_{\mathbf{k}}})
\right]^{2} \nonumber \\
& & \times N_q (N_k+1) 
\delta(\epsilon_f-\epsilon_i-\hbar \omega_{\mathbf{q}} + \hbar \omega_{\mathbf{k}}) \, ,
\label{eq:lifetime}
\end{eqnarray}
where $N_{\mathbf{q}}=1/\left[\exp(\hbar \omega_{\mathbf{q}} / k_B T)-1 \right]$
is the number of phonons at the given temperature $T$. Only long-wave acoustic phonons
are involved in the process, where
$\omega_{\mathbf{q}} = c_{\nu} |\mathbf{q}|$, and $c_\nu$ is the sound speed for the
$\nu=\text{LA}$ (longitudinal acoustic phonon) and TA (transverse acoustic phonon)
modes. The $'$ in the equation indicates that the summation includes all the (intermediate)
states except for the initial and final states. The hole-phonon interaction
matrix elements $M_{\mathbf{q}}^{is}$ are given by:
\begin{equation}
M_{\mathbf{q}}^{is}= \alpha_{\nu}(\mathbf{q}) \langle\psi_{i}|e^{i\mathbf{q}\cdot\mathbf{r}}|\psi_{s} \rangle \, ,
\label{eq:Mq}
\end{equation}
where $|\psi_{i}\rangle$ and $|\psi_{s}\rangle$ are the initial and intermediate state wave functions,
respectively. $\alpha_{\nu}(\mathbf{q})$ is the hole-phonon coupling strength. We have considered
three hole-phonon 
interaction mechanisms in the QDs~\cite{cheng04,wei12}: hole-acoustic-phonon interaction due to (i) the deformation
potential ($\nu$ = LADP), (ii) the piezoelectric field for the longitudinal mode ($\nu$ = LAPZ), and
(iii) the piezoelectric field for the transverse mode ($ \nu $ = TAPZ). $\alpha_{\nu}(\mathbf{q})$ and
other parameters used in the calculations can be found in Ref.~\onlinecite{wei12}. The overall spin
relaxation time $1/T_1^h=\sum_{\nu} 1/\tau_{\nu}$.

To calculate $T_1^h$, we use the atomistic EPM to obtain high-quality hole energy levels and wave
functions.~\cite{wei12}
We simulate a lens-shaped In$_{1-x}$Ga$_x$As/GaAs QDs embedded in a cubic GaAs
matrix, containing 60$\times$60$\times$60 GaAs 8-atom unit cells, as  
illustrated in the Fig.~\ref{fig:QDs}(b). 
The dot is grown along the [001] direction, on the top of a monolayer wetting layer.
We first obtain optimized atomic positions $\{\mathbf{R}_{n,\alpha}\}$
($\alpha$-th atom at the site $n$) 
by minimizing the total strain energy of the system (matrix+QDs) 
via valence force field (VFF) method.~\cite{keating66}
The hole energy levels and wave functions are obtained by solving the following Schr\"{o}dinger equation,
\begin{equation}
\left[ -\frac{1}{2}\nabla^2 + V_{\text{epm}}(\mathbf{r}) +
  V_{\text{ef}}(\mathbf{r}) 
+ \frac{1}{2}g\mu_{B} B_z \sigma_z
\right] \psi_i (\mathbf{r})   = \epsilon_i \psi_i (\mathbf{r}) \, ,
\end{equation}
where $V_{\text{epm}}(\mathbf{r})  = \sum_{n, \alpha} \hat{v}_{\alpha} (\mathbf{r} - \mathbf{R}_{n,\alpha})
+ V_{\text{SO}}$ is the total screened electron-ion potential, 
including the superposition of all atomistic pseudopotentials
$\hat{v}_{\alpha} (\mathbf{r})$ and 
the non-local spin-orbit potential $V_{\text{SO}}$.~\cite{wei12} 
This method naturally
includes the Rashba and Dresselhaus SOC in a ``first-principles''
manner. $V_{\text{ef}}$ is the external potential due to the applied electric field, 
along the growth direction [see Fig.~\ref{fig:QDs}(b)].
$E>$0 ($E<$0) corresponds to that the electric field points to the [001] ([00$\bar{1}$])
direction. We also applied an extremely small magnetic filed ($B_z = 1$ mT) along the growth direction to split the
spin-up and spin-down states, where $\sigma_z$ is the Pauli matrix
and $g$=2 is the Lande $g$ factor. 
The spin-up and spin-down energy difference caused by the magnetic
field is negligible ($\sim 0.12$ $\mu$eV).

The Schr\"{o}dinger equation is solved by the linear combination of bulk bands (LCBB) method.~\cite{wang99b}
We use eight bands (including spin) for the hole in the calculation, which
takes both the inter-valence-band coupling and the valence-conduction band
coupling into account. A 6$\times$6$\times$16
k-mesh converges the energy and wave functions very
well.~\cite{wei12,williamson00} 
Due to the SOC, the wave functions are spin mixed, i.e. $|\psi_i\rangle =
\alpha |\uparrow\rangle + \beta |\downarrow\rangle$. 
We regard $|\psi_i\rangle$ as a spin up (down) state if $\alpha > \beta$
($\alpha < \beta$).
To calculate $T_1^h$,
we sum over 40 intermediate states (including spin), which
converges the results within 0.1 ms.~\cite{wei12}

\begin{figure}
\includegraphics[width=0.5\textwidth]{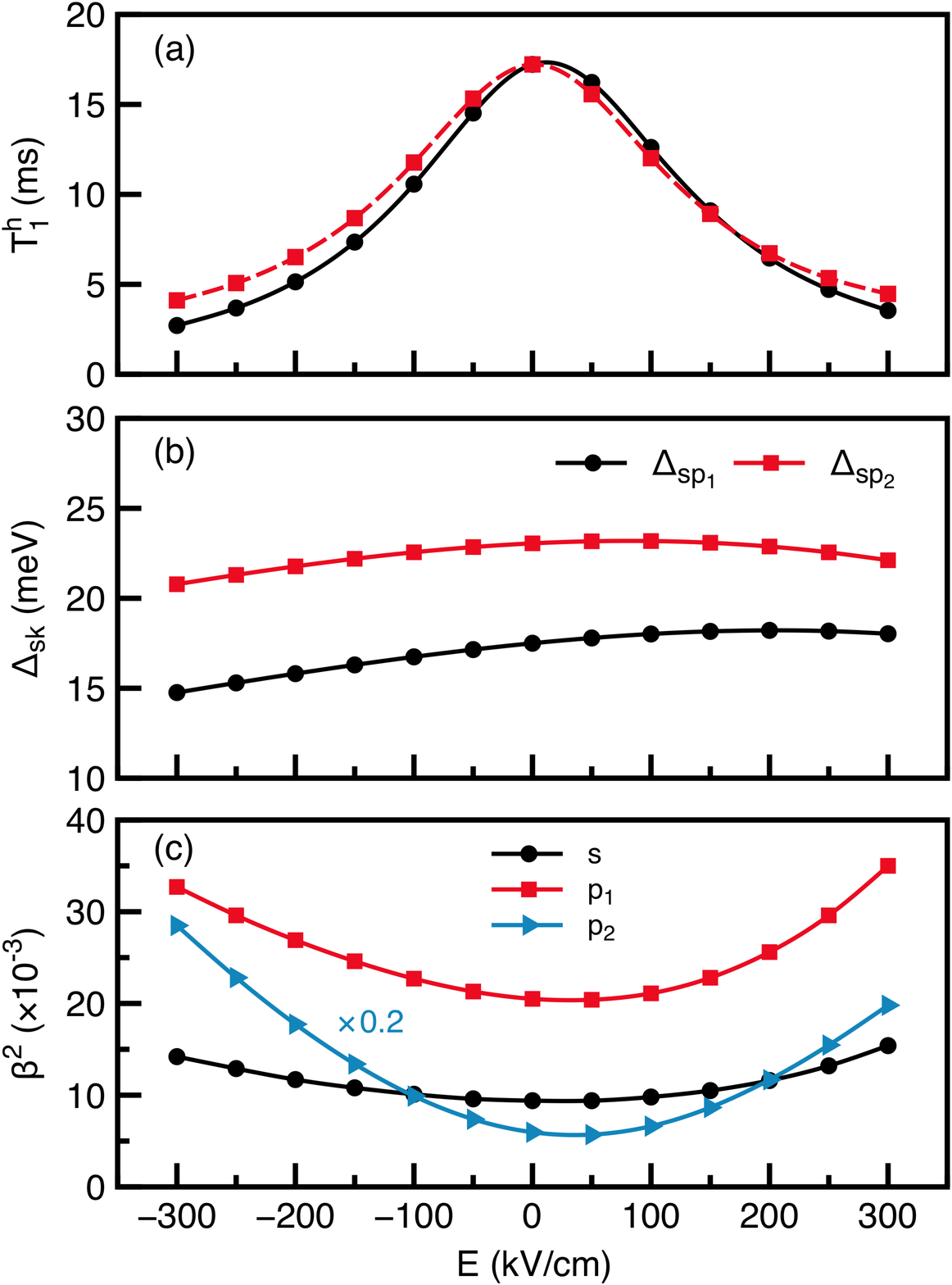}
% Here is how to import EPS art
\caption{(Color online) (a) Black solid line: 
the hole relaxation time $T_1^h$ as a function of $E$ in
a pure InAs/GaAs QD, with $b$=20 nm, and $h$=1.5 nm. 
Red dashed line: same as above, but the hole energy levels are artificially
fixed to those of $E=0$ kV/cm.
(b) The energy spacing between the $s$ level and $p$ levels
as functions of $E$.
(c) The spin mixture parameters $\beta^2$ as functions of $E$
for the $s$ (black), $p_1$ (red) and $p_2$ (blue) levels.}
\label{fig:fig2}
\end{figure}

Figure~\ref{fig:fig2}(a) depicts 
the hole relaxation time $T_1^h$ (black solid line) at 4.2 K as a
function of electric field ($E$) for a pure InAs/GaAs QDs
with base diameter $b$=20 nm and height $h$=1.5 nm. We apply the
electric field between $-300$ and $300$ kV/cm, where the hole can still be
trapped in the QDs. When no electric filed is applied ($E=0$), we
find the hole spin relaxation time $T_1^h$=17.2 ms. For $E>0$, $T_1^h$
decreases rapidly with the increasing $E$. At $E=300$ kV/cm,
$T_1^h$ decreases to 3.5 ms.
For $E<0$, with the increasing of $|E|$,
$T_1^h$ decreases to 2.7 ms at $E=-300$ kV/cm.
The short spin decay time may be useful 
in some cases, for example, fast spin initialization. 
The longest $T_1^h$ is approximately 17.4 ms 
at $E$=17.2 kV/cm. 
%The hole spin relaxation show asymmetry with the respect of $E$, 
%which is the result of symmetry breaking of QDs along [001] direction. 

As discussed in our previous work,~\cite{wei12} 
$T_1^h$ is determined by two factors: one is the energy 
difference $\Delta_{sk}$ between the lowest level ($s$) and the intermediate
levels ($k$). In this case, smaller $\Delta_{sk}$ can fasten the relaxation process.
The other is minor spin component $\beta$, 
which reflects the
spin-up and spin-down mixture due to SOC. And in this case, lager $\beta$ leads to a smaller $T_1^h$.
We find that the electric field can tune the energy spacing between the $s$ and $p_1$ and $p_2$ level
by approximately 2$\sim$4 meV, as shown in Fig.~\ref{fig:fig2}(b).
Figure~\ref{fig:fig2}(c) shows the
$\beta^2$ of the $s$ (black line), $p_1$ (red line) and $p_2$ (blue line) 
states as functions of $E$. We find that the electric field can significantly
change the spin mixture of the wave functions, due to the change of SOC by
electric field.
To determine the main mechanism that causes the change of $T_1^h$,
we artificially fix the hole energy levels at different applied electric
fields to the ones at $E$=0 kV/cm and recalculate $T_1^h$. The results are
shown in Fig.~\ref{fig:fig2}(a) in the dashed red line, which are
rather close to the results using the electric field dependent energy levels. 
This clearly suggests that spin mixture tuned by electric field plays a major 
rule in tuning $T_1^h$.

\begin{figure}
\includegraphics[width=0.5\textwidth]{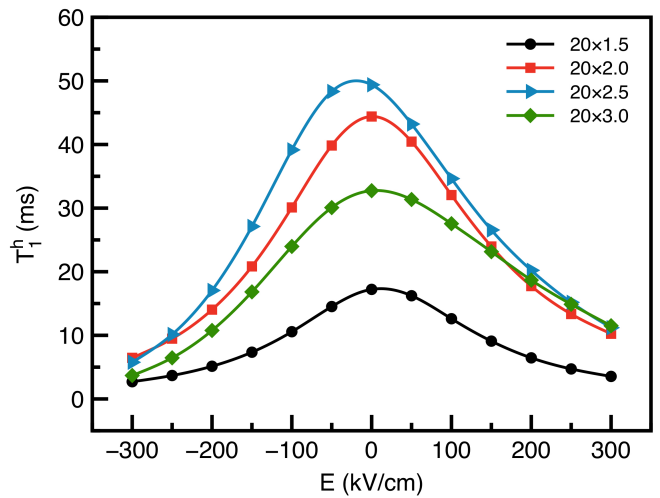}% Here is how to import EPS art
\caption{\label{fig:t1_pure} (Color online) The hole spin relaxation times as
  functions of electric field in InAs/GaAs QDs, with dot diameter $b$=20
  nm,  and height $h$=1.5, 2.0, 2.5 and 3.0 nm.}

\end{figure}

We further calculate $T_1^h$ at 4.2 K for 
QDs of different geometries. 
Figure~\ref{fig:t1_pure} depicts $T_1^h$ as a function of $E$ in pure
lens-shaped InAs/GaAs QDs with fixed base diameter $b$=20 nm whereas the dot
hight $h$ varies from 1.5 nm to 3.0 nm. 
For all QDs, the hole spin relaxation times are tuned by electric field in 
a very similar way. 
The spin relaxation time tends to decrease with $|E|$.
However, for the flat QDs, the tuning of $T_1^h$ by electric 
field is rather symmetric, whereas for taller dots the tuning
becomes more asymmetric, because the geometry of dots themselves
become more asymmetric. 
%In the taller QDs with h=3.5 nm, when electric field is applied along
%$[00\bar{1}]$ direction, $T_1$ increases with decreasing $E$. When electric
%field is applied along $[001]$ direction, $T_1$ increases until about 100
%kV/cm and then decreases.
In all cases, the hole spin relaxation time can be 
tuned by more than one order of magnitude.
For example, $T_1^h(-300)/T_1^h(0) = 0.11$ in the 20 nm$\times$3.0 nm QDs. 

\begin{figure}
\includegraphics[width=0.5\textwidth]{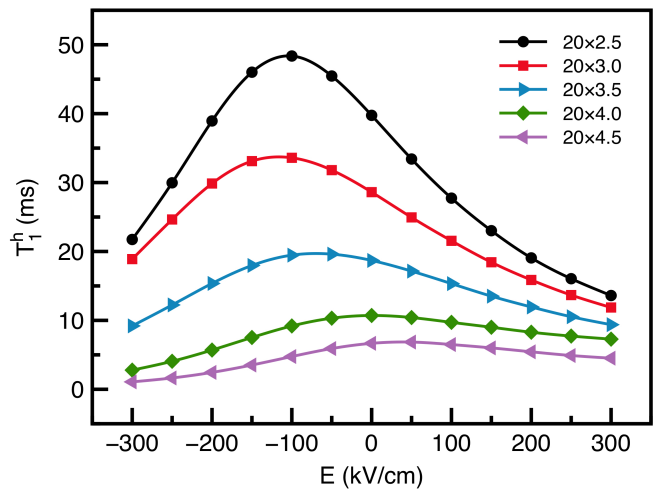}% Here is how to import EPS art
\caption{\label{fig:t1_alloy_Ga020} (Color online) The hole spin relaxation
  times as functions of electric field in In$_{0.8}$Ga$_{0.2}$As/GaAs QDs,
  with dot diameter $b$=20 nm, and height $h$=2.5, 3.0, 3.5, 4.0 and 4.5 nm.
%The insert shows the electric field $E_M$ that has the longest $T_1^h$ for
%different dot heights.
}
\end{figure}

We find similar results for alloy In$_{0.8}$Ga$_{0.2}$As/GaAs QDs. We
calculate $T_1^h$ for the dots with base diameter $b$=20 nm, 
and the dot height $h$ varying from 2.5 nm to 4.5 nm.
The results are shown in Fig.~\ref{fig:t1_alloy_Ga020}. 
Remarkably, the hole spin relaxation time can be significantly prolonged 
by the electric field in the alloy dots.
Take the 20 nm$\times$2.5 nm QDs (black line in
Fig.~\ref{fig:t1_alloy_Ga020}) as an example, 
$T_1^h$ first increases with the negative 
electric field and reaches to a
maximum value 48 ms at about $E_M=-100$ kV/cm. It starts to decrease 
when the electric field further increases.
When a positive electric field is applied, $T_1^h$ decreases monotonically.
The other dots show similar behaviors.
However, with the increasing of the dot height, the $E_M$ which has the longest 
$T_1^h$ generally shifts to the more positive direction, as shown in 
Fig.~\ref{fig:t1_alloy_Ga020}.

To conclude, we have investigated the tuning of hole spin relaxation
in single self-assembled In$_{1-x}$Ga$_{x}$As/GaAs QDs by electric field
using an atomistic empirical pseudopotential method.
We find that the electric filed can significantly 
increase or decrease the hole spin relaxation time in QDs, which
provides an effective way to tune the hole spin relaxation time that may be
useful for future device applications.

LH acknowledges support from the Chinese National
Fundamental Research Program 2011CB921200
and National Natural Science Funds for Distinguished Young Scholars.

%\bibliographystyle{aipnum4-1}
%\bibliography{MyDotBiblio4,DotBiblio}% Produces the bibliography via BibTeX.

%merlin.mbs aipnum4-1.bst 2010-07-25 4.21a (PWD, AO, DPC) hacked
%Control: key (0)
%Control: author (8) initials jnrlst
%Control: editor formatted (1) identically to author
%Control: production of article title (-1) disabled
%Control: page (0) single
%Control: year (1) truncated
%Control: production of eprint (0) enabled
%

\end{document}